\documentclass[prl,reprint]{revtex4-2}
\usepackage{graphicx,color}
\usepackage{amsmath,ulem}
\usepackage[hidelinks]{hyperref}
\usepackage{bm}

\begin{document}
\title{The supercurrent diode effect and nonreciprocal paraconductivity due to the chiral structure of nanotubes }
\author{James Jun He$ ^{1,2,*} $}
\author{Yukio Tanaka$ ^3 $}
\author{Naoto Nagaosa$ ^{2,4} $}
\affiliation{
	$ ^1 $International Center for Quantum Design of Functional Materials (ICQD), 
	Hefei National Laboratory for Physical Sciences at Microscale, 
	University of Science and Technology of China, Hefei, Anhui 230026, China \\
	$ ^2 $Center for Emergent Matter Science (CEMS), RIKEN, Wako, Saitama, 351-0198, Japan\\	
	$ ^3 $Department of Applied Physics, Nagoya University, Nagoya, 464-8603, Japan\\
	$ ^4 $Department of Applied Physics, University of Tokyo, Tokyo, 113-8656, Japan \\
	*Corresponding author. Email: jun\_he@ustc.edu.cn
}

\begin{abstract}
	The research interest in the supercurrent diode effect (SDE) has been growing. It has been found in various kinds of systems, in a large part of which it may be understood by combining spin-orbit coupling and Zeeman field. Here, we show that there exists another mechanism of generating SDE in chiral nanotubes that trap magnetic fluxes, without spin-orbit coupling or Zeeman field. We further show that the same generalized Ginzburg-Landau theory leads to nonreciprocal paraconductivity (NPC) near the transition temperature. The main features of both the SDE and the NPC are revealed by their parameter dependence. Our study suggests a new kind of platforms to explore nonreciprocal properties of superconducting materials. It also provides a theoretical link between the SDE and the NPC, which were often studied separately. 
\end{abstract}
\maketitle 

\paragraph{\bf INTRODUCTION\\}

Nonreciprocal transport properties \cite{Tokura} near or inside the superconducting phase of electronic systems have been attracting a lot of research attention recently.
It may manifest itself in nonreciprocal paraconductivity (NPC) \cite{Wakatsuki1,Wakatsuki2,Hoshino,Masuko,Iwasa1,Iwasa2,Yasuda} or in so-called supercurrent diode effect (SDE) \cite{Ando,JPHu}.

In superconductors (SCs) or Josephson junctions with broken inversion ($ \mathcal{P} $) and time-reversal ($ \mathcal{T} $) symmetries, the critical currents along opposite directions, $ J_{c\pm} $, may  be unequal, leading to the SDE. This effect has been found  in various experimental systems \cite{Ando, Baumgartner, JXLin, Jiang,Pal,Bauriedl,Ali,Baumgartner2,Narita}, part of which may be understood by combining spin-orbit coupling (SOC) and Zeeman field \cite{Noah,JJH,Daido} which break $ \mathcal{P}  $ and $ \mathcal{T} $, respectively. 
The SOC-Zeeman mechanism also works in one-dimension \cite{XJLiu,Legg} and in systems with disorders \cite{Ilic}.
There also exist theories that consider symmetry breakings by internal magnetic \cite{JiangK,Scammell,Halterman,Kokkeler,Karabassov}, electric \cite{Zhai,JYChen} or valley \cite{YMXie} orders, finite momentum pairing \cite{Davydova,YMXie2}, unconventional superconductivity \cite{Zinkl,Tanaka}, etc.
However, systems with magnetic orders may be understood in a way similar to those under Zeeman fields, and superconductors with ferroelectric or valley orders, finite momentum pairing, or  $ \mathcal{T} $-breaking are not conveniently found in nature.
Thus, it remains an open question whether there exist a new mechanism to generate the SDE in state-of-the-art experimental systems. Finding such a mechanism shall greatly enrich the choice of platforms to investigate the SDE and promote the research in this direction. 

While the SDE is a manifestation of a nonreciprocal SC below its transition temperature $ T_c $, the nonreciprocity can also be seen slightly above $ T_c $, where Cooper pairs start to form but coherent superconductivity is not reached yet. In this regime, the trend of forming Cooper pairs makes a large contribution to the conductivity, which is called the paraconductivity \cite{AL, Maki}. In systems where $ \mathcal{P}$  and $ \mathcal{T} $  are broken, the paraconductivity in opposite directions may differ significantly, leading to the NPC. Although nonreciprocal conductance may also exist in the normal state at $ T \gg T_c $, this effect can be enhanced by several orders of magnitude as the temperature approaches $ T_c $ \cite{Wakatsuki2}. Theories have shown that the NPC can also originate from a combination of SOC and Zeeman field \cite{Wakatsuki1,Wakatsuki2}. Despite the similarity in the conditions to realize SDE and the NPC, current theories have not discussed the two in the same framework to the best of our knowledge.


Here, we show that both the SDE and the NPC exist in a chiral nanotube under a magnetic field along its axial direction, and they can be obtained with the same generalized Ginzburg-Landau theory. The inversion symmetry is broken by the chiral structure of the nanotube without any SOC, and the magnetic field plays its role through the orbital effect, i.e., Aharonov-Bohm effect, instead of the Zeeman coupling. The resulting nonreciprocal signals strongly depend on the magnetic flux, the nanotube radius, and the chiral angle. There exist a periodicity in the magnetic flux through the tube, similar to the Little-Parks oscillation \cite{LittleParks}, as well as a periodicity in the chiral angle. The interplay of the magnetic flux and the chiral structure is the origin of both the SDE and the NPC.  
\\

\paragraph{\bf RESULTS\\}
\paragraph{\bf Chiral nanotubes near $ T_c $ \\}

\begin{figure}
	\includegraphics[width=0.80\linewidth]{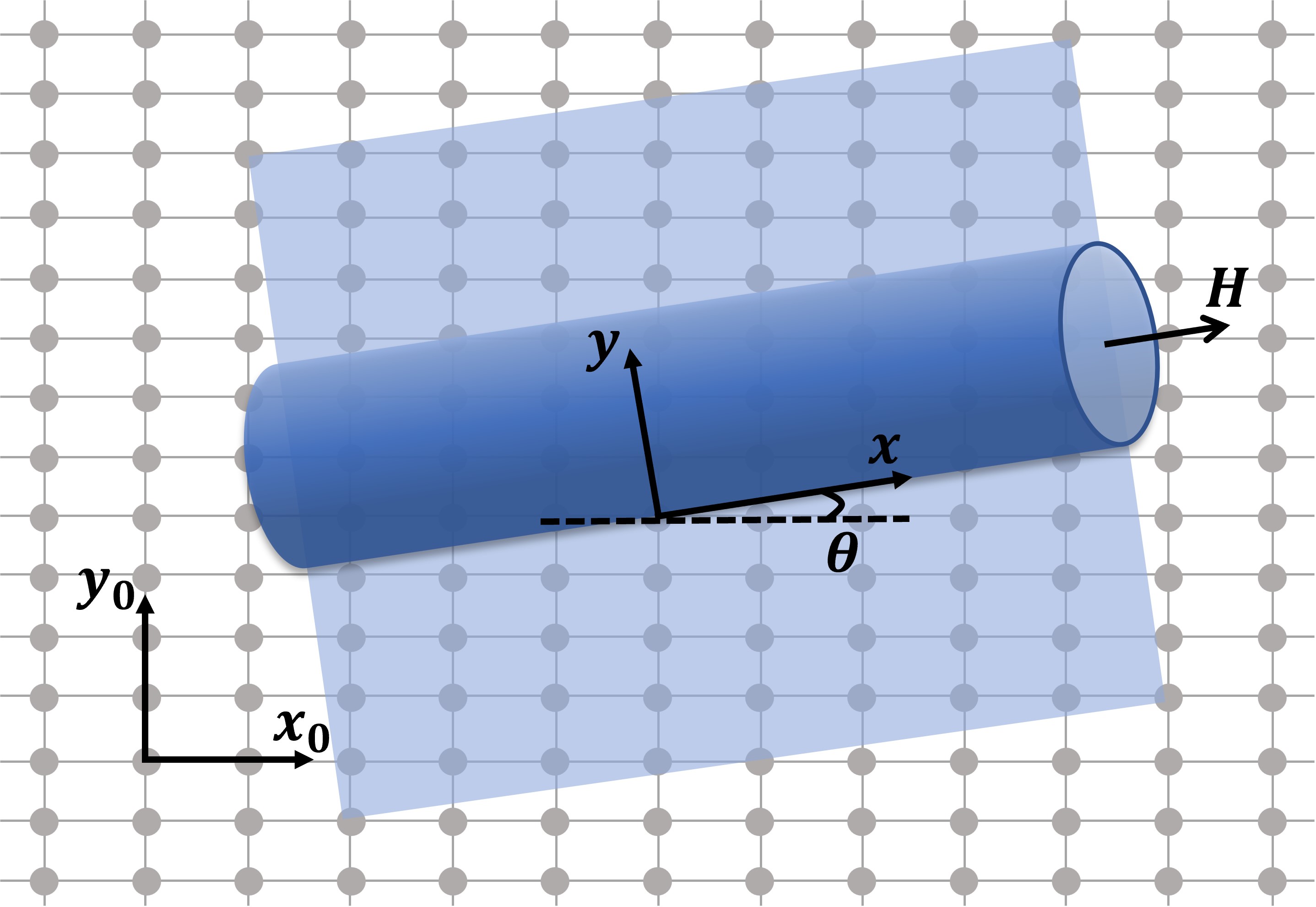}
	\caption{A schematic of a chiral nanotube formed by rolling a two-dimensional sheet. The two coordinate systems, $ (x_0, y_0) $ and $ (x,y) $, are connected by a rotation of the chiral angle $ \theta $. A magnetic field $ \bm H $ is applied along the tube to generate nonreciprocal effects.} 
	\label{fig:tube}
\end{figure}

A nanotube near its superconductivity transition temperature $ T_c $ may be described by the following free energy,                                       
\begin{align}
	F= \int d^2\bm r  \psi^*(\bm r) [\alpha  + \xi(\hat{\bm p}) + \frac{\beta}{2} |\psi(\bm r)|^2 ] \psi(\bm r),
	\label{eq:F}
\end{align}
where $ \alpha \sim T-T_c $ and $ \beta  $ are the conventional Ginzburg-Landau parameters. The displacement vector $ \bm r = (x,y) $ is defined so that the nanotube aligns alone the $ x $-direction and the transverse coordinate $ y $ circulates around the tube, as illustrated in FIG. \ref{fig:tube}. The term $ \xi( \hat{\bm p}) = \sum_{ij} \xi_{ij} \hat{p}_x^i \hat{p}_y^j $ is the kinetic energy of a Cooper pair. Apparently, a periodic boundary condition should be applied along the $ y $-direction. 
The momentum operator is $ \hat{\bm p} = -i \hbar  \nabla_{\bm r} + 2e\bm A(\bm r) $. 
Considering a uniform magnetic field applied along the $ x $-direction, i.e  $ \bm H = H_x  \hat{\bm x}$, and assuming the nanotube wall thickness to be negligible, the vector potential becomes $ \bm A = \frac{\phi }{2\pi R}\hat{\bm y} $, where $ \phi=\pi R^2 H_x $ is the magnetic flux through the nanotube and $ R $ is its radius. This is equivalent to a boundary condition $ \psi(\bm r) = \psi(\bm r + 2\pi R  \hat{\bm y}) \exp\{-2\pi i\phi/\phi_0\}$,  $ \phi_0 = h/2e $ being the magnetic flux quantum. 

A Fourier transformation (taking into account the magnetic flux)
leads to the following equivalent form of Eq. (\ref{eq:F}),
\begin{align}
	F= &2\pi R \sum_{n} \int d q [\alpha + \xi\left( \bm p  \right) + \frac{\beta}{2} (2\pi R)^2 |\psi_{n}|^2 ]  |\psi_{n}|^2 ,
	\label{eq:F2}
\end{align}
where $ q $ is the wavenumber along the tube and $ \bm p =  ( \hbar q, [n- \phi/\phi_0]\hbar/R )$. The integer $ n $ labels the transverse Fourier components. It is quantized due to the small circumference of the tube.
We have neglected the coupling between different $ \bm q $-components in the $ |\psi|^4 $ term, which does not affect the results of this study. 
It is clear from Eq. (\ref{eq:F2}) that $ F$ is a periodic function of $ \phi $, leading to the Little-Parks oscillation, as will be seen later. 

The chiral structure of the nanotube is reflected in the functional form of $ \xi( \bm p) $. To see that, imagine a nano-ribbon obtained by cutting and flattening the nanotube. When the local continuous rotational symmetry ($ \mathcal{C}_\infty $) of this ribbon is reduced a discrete $ \mathcal{C}_n $, a chiral nanotube can be obtained if the rolling direction mismatch all the high-symmetry directions. For simplicity, we consider here a system with $ \mathcal{C}_2 $ and the kinetic term may be written as (up to the 4-th order in the momentum)
\begin{align}
	\xi(\bm p _0) = &\frac{|\bm p_0|^2}{2m_0} + \frac{|\bm p_0|^4}{4m_0^2 \zeta_0 } + \frac{p_{x0}^2-p_{y0}^2}{2m_1}  +  \frac{ ( p_{x0}^2-p_{y0}^2)^2}{4m_1^2 \zeta_1}\notag\\
	&+\frac{p_{x0}^2( p_{x0}^2-3 p_{y0}^2)+p_{y0}^2( p_{y0}^2-3 p_{x0}^2)}{4m_2^2 \zeta_2}
	\label{eq:xi0}
\end{align}
where $ \bm p_0 $ is defined in a coordinate system whose axes align with the high-symmetry directions. It is generally different from that of $ \bm p $ defined in the previous coordinate system whose $ x $-axis is along the nanotube. They are connected by a rotation of the chiral angle $ \theta $, as shown in FIG. \ref{fig:tube}. The first two terms in Eq. (\ref{eq:xi0}) preserves $ \mathcal{C}_\infty $ while the third term reduces it to $ \mathcal{C}_2 $.  
Note that $ m_1>m_0 $ must hold for the mass along arbitrary direction to be positive. 
The last two terms are $ \mathcal{C}_4 $  symmetric. The inclusion of quartic terms is necessary to reveal the nonreciprocal properties, similar to the case where such an effect is caused by magnetochiral anisotropy \cite{Wakatsuki1,Wakatsuki2,JJH,Daido}.

Equation (\ref{eq:xi0}) can be rewritten as
\begin{align}
\xi(\bm p)=\frac{p_x^2}{2m_x}+\frac{p_y^2}{2m_y}+\frac{p_x p_y}{m_{xy}} + \sum_{n=0}^4 \kappa_n p_x^n p_y^{4-n}
\label{eq:xi}
\end{align}
with $ m_x, m_y, m_{xy} $ and $ \kappa_n $ being functions (see Materials and Methods) of the original parameters in Eq. (\ref{eq:xi0}).
To see how a chiral nanotube breaks $ \mathcal{P} $, note that $ p_y = (n_y-\phi/\phi_0)\hbar/R $ is defined along a circular coordinate and behaves as angular momentum (rather than the usual momentum in a flat space). It remains unchanged under $ \mathcal{P} $ operation, consistent with the symmetry property of the magnetic flux $ \phi $ which should not change under spatial inversion. As a result, the nanotube geometry leads to the symmetry operation $ (p_x,p_y) \xrightarrow{\mathcal{P}}  (-p_x, p_y) $, and thus the $ p_x $-odd terms in Eq. (\ref{eq:xi}) break $ \mathcal{P} $. 

The supercurrent is
\begin{align}
	J_x &
	=  -2e \int dy \psi^*(\bm  r) \frac{d \xi}{d \hat{p}_x} \psi(\bm r)   
	\\
	&= -2e \sum_{n} \frac{2\pi R}{L} \int dq \frac{\partial \xi(\bm p)}{\partial p_x} |\psi_{n}(q)|^2,
	\label{eq:Jx0}
\end{align}
where $ L \rightarrow \infty $ is the length of the nanotube. With Eqs. (\ref{eq:F2}), (\ref{eq:xi}) and (\ref{eq:Jx0}), we study the SDE when $ T < T_c $ and the NPC when $ T>T_c $ in the following. \\

\paragraph{\bf Supercurrent diode effect\\}

When a supercurrent passes through the nanotube, the Cooper pairs acquire a momentum $ \bm p$ and a kinetic energy $ \xi(\bm p) $. The order parameter is determined by the Ginzburg-Landau equation as 
\begin{align}
	|\psi_n(q)|^2 = \frac{|\alpha |}{\beta (2\pi R)^2 } \left(1-\frac{\xi(\bm p)}{|\alpha |}\right)
\end{align}
and the supercurrent is 
\begin{align}
    J_x (n,q) &=  \frac{-2e R}{L^2} \frac{|\alpha |}{\beta R^2}  \left(1-\frac{\xi(\bm p)}{|\alpha |}\right) \frac{\partial \xi(\bm p)}{ \partial p_x}  .
    \label{eq:Jx1}
\end{align}
Note that $ \alpha <0 $ since $ T<T_c $. The critical currents $ J_{c\pm} $ are the absolute values of the maximum and minimum, respectively, of $ J_x(n,q) $ as $ n $ and $ q $ are varied. 

\begin{figure}
	\includegraphics[width=0.38\textwidth]{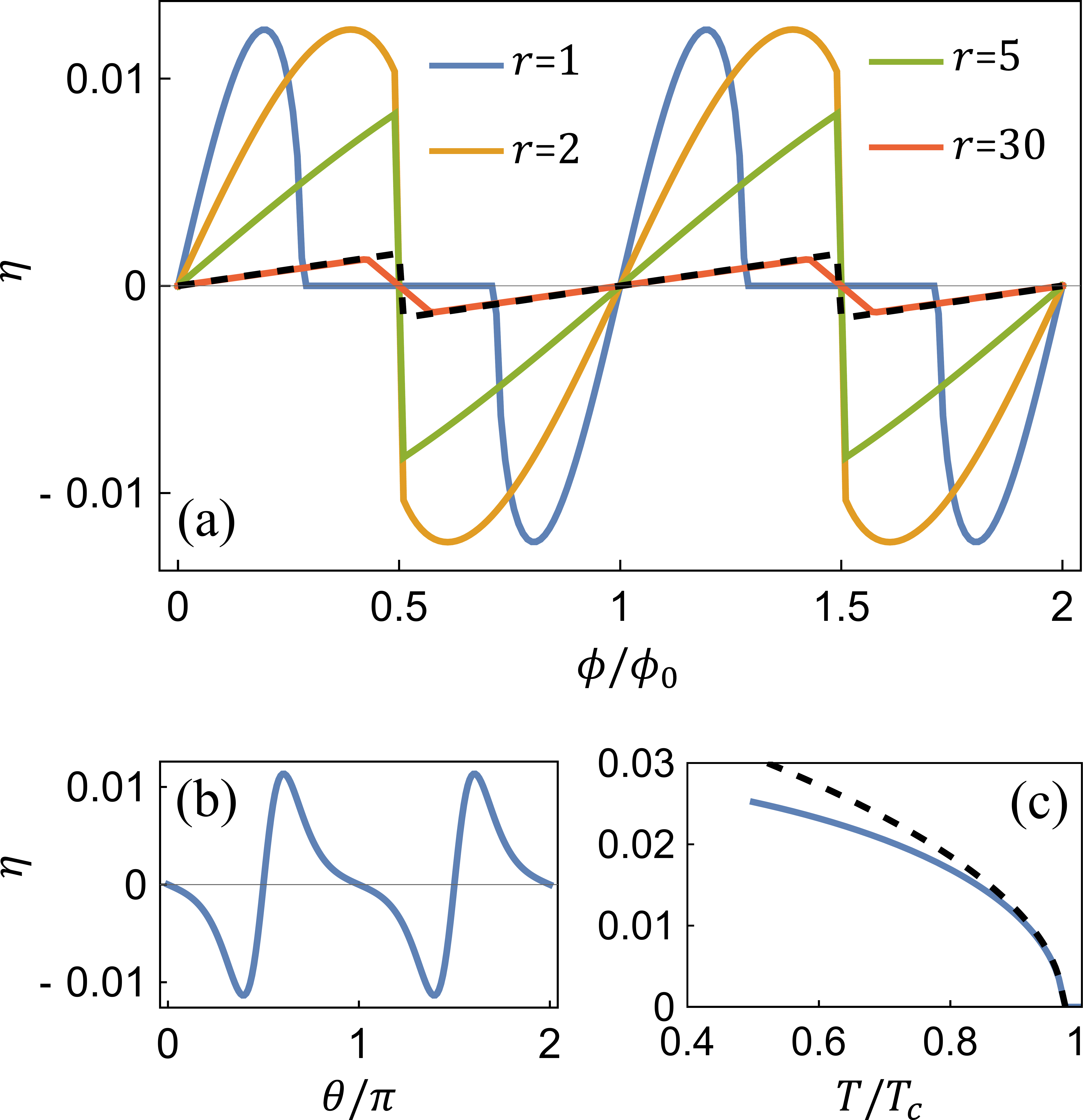}
	\caption{The diode efficiency, $ \eta = (J_{c+}-J_{c-})/(J_{c+}+J_{c-}) $, 
	obtained by numerically solving for the critical currents $ J_{c\pm} $ with Eq. (\ref{eq:Jx1}). (a) The dependence on the magnetic flux ($ \phi_0=h/2e $ is the flux quantum). The solid curves are for various values of the nanotube radius $ R $, normalized so that $ r=R/l_0 $, where $ l_0 = \hbar/\sqrt{2m_0 T_c} $. The dashed curve is the approximate result given by Eq. (\ref{eq:eta}) with $ r=30 $. (b) Dependence on the angle $ \theta $ which corresponds to the chiral structure of the nanotube. (c) The temperature dependence. The parameters are
	$  m_0=1, m_1=2, \zeta_2 m_2 \rightarrow \infty,  \zeta_0/T_c=10 , \zeta_1/T_c=20, r=2 , \theta=0.6\pi , \phi/\phi_0=0.3 $ and $ T/T_c=0.9$ for all the results unless specified otherwise. }
	\label{fig:eta}
\end{figure}

For general parameters, $ J_{c\pm} $ can be determined numerically and the resulting diode efficiency, $ \eta\equiv \frac{I_{c+}-I_{c-}}{I_{c+}+I_{c-}} $, is shown in FIG. \ref{fig:eta} as functions of the magnetic flux $ \phi $, the angle $ \theta $ and the temperature, respectively.
FIG. \ref{fig:eta}(a) shows a periodicity in $ \phi $, similar to the Little-Parks oscillation. 
Different curves are for various values of the ratio $ r=R/l_0 $, with $ R $ being the radius of the nanotube and $ l_0=\hbar/\sqrt{2m_0 T_c} $. 
When $ r $ is small and $ \phi/\phi_0 $ is close to a half-integer, the transverse momentum, $ p_y = (n-\phi/\phi_0) \hbar/R \approx \hbar/(2r l_0)  $, costs so high a kinetic energy $ \xi(\bm p) $ that it kills the superconductivity (i.e., $ \psi_n \rightarrow 0$), leading to vanishing $J_{c\pm}$. 
We define $ \eta $ in this case to be zero, resulting in the curve with $ r=1 $ in FIG. \ref{fig:eta} (a). 
As $ r $ increases, $ J_{c\pm} $ becomes nonzero for arbitrary magnetic flux and discontinuities occur as $ \phi/\phi_0 $ changes across half-integers, which originates from the quantization of the transverse index $ n $ in Eq. (\ref{eq:Jx1}). When $ r \gg 1 $, discontinuities disappear while non-smooth kinks remain and $ |\eta| $ decreases. 
From FIG. \ref{fig:eta}(b), one finds that $ \eta $ vanishes whenever $ \theta $ becomes a multiple of $ \pi/2 $. This is expected because the nanotubes in these cases are not chiral and the inversion symmetry is preserved, forbidding the SDE. As $ \theta /\pi $ deviate from  half-integers, $ |\eta| $  increases sharply and extreme values of $ \eta $ are reached quickly. Note that the positions of the extreme points depend on the ratio $ m_0/ m_1 $, which measures the strength (and the sign) of inversion symmetry breaking. The temperature dependence has the usual feature $ \eta\sim \sqrt{T_c-T} $, as shown in FIG. \ref{fig:eta} (c).

It is helpful to obtain the analytical form of $ \eta $, which is possible when $ \zeta_{0,1,2} \gg T_c $ and thus the terms with $ \kappa_n $ in Eq.(\ref{eq:xi}) can be treated as perturbations. We also assume $ r $ to be small, and then
varying the transverse quantum number $ n $ costs so much energy that $ J_{c\pm} $ are obtained with a fixed $ n $ in Eq. (\ref{eq:Jx1}). Under these  conditions, the diode efficiency is
\begin{align}
	\eta
	= 
	\frac{-4 }{\sqrt{3}}&\left(4 \kappa _0  \frac{ m_x^2 }{ m_{xy}}+\kappa _1 m_x\right) m_0 T_c   
	\notag \\ 
	&\times  b\sqrt{\frac{|\alpha| }{T_c} \frac{m_x}{m_0}-b^2 \left(\frac{m_x}{m_y}-\frac{m_x^2}{m_{xy}^2}\right)} ,
	\label{eq:eta}
\end{align}
where $ b = \phi/\phi_0 - [\phi/\phi_0] $ ($ [x] $ denotes the integer closest to $ x $).  
From Eq. (\ref{eq:eta}) it becomes clear that either $ m_{xy}^{-1} $ or $ \kappa_1 $ must be nonzero to achieve the SDE. The requirement, combined with Eqs. (\ref{eq:mxy}) and (\ref{eq:kappa1}), becomes $ m_1^{-1}\neq 0 $ and $ \sin 2\theta \neq 0 $, which is just equal to requiring the nanotube to have a chiral structure. 
When the magnetic filed $ H_x $ is small, $ \eta $ is linear in $ H_x $ (note that $ \phi=\pi R^2 H_x $). As the magnetic flux increases, the expression under the square root becomes negative for small $ |\alpha| $ since $ (\frac{m_x}{m_y}-\frac{m_x^2}{m_{xy}^2})  $ is positive definite. This results in a decrease of 
the transition temperature to $ T_c' $ with $ \delta T_c = T_c-T_c' \sim  b^2 (\frac{m_x}{m_y}-\frac{m_x^2}{m_{xy}^2}) \frac{m_0}{m_x} $. And the temperature dependence of Eq. (\ref{eq:eta}) may be written as $ \eta \sim \sqrt{T_c'-T} $. 
A substitution of Eqs. (\ref{eq:mx})$ - $(\ref{eq:kappa2}) leads to the dashed curves in FIG. \ref{fig:eta} (a) and (c), which show great agreement with previous numerical results except two situations,  (i) $ r\gg 1 $ and $ \phi/\phi_0 $ is close to a half integer and (ii) The temperature is far below $ T_c $. In both situations, the assumption that $ J_{c\pm} $ can be obtained with the same index $ n $ in Eq. (\ref{eq:Jx1}) no longer holds.

The differences in the SDE between chiral nanotube SCs and previously studied spin-orbit coupled SCs \cite{Noah,JJH,Daido} is clear now. 
The diode efficiency here is controlled by the nanotube diameter and the chiral angle, while it is determined by the SOC strength in spin-orbit coupled SCs. The sign change of $ \eta $ happens  in both kinds of systems as the magnetic field is tuned. However, the origins are rather different. 
In SOC SCs, $ \eta $ changes sign due to the higher-order (in momentum and in field strength)  terms in the kinetic energy of the Cooper pairs. 
Here, it is because the transverse index $ n $ corresponding to the critical currents $ J_{c\pm} $ is shifted. The sign of $ \eta  $ changes exactly at $ b = 1/2$  here (i.e. when the number of flux quanta is a half-integer) while the sign-flipping field-strength in SOC SCs depends on multiple system parameters. 
\\

\paragraph{\bf Nonreciprocal paraconductivity \\ }

The nonreciprocity of superconducting materials manifests itself not only in the SDE when $ T-T_c<0 $, but also in the NPC when $ T_c\gg  T-T_c >0 $. In the latter case, although the average order parameter vanishes, its quantum fluctuations induce a significant contribution to the conductance, resulting in a drop of resistance  above $ T_c $ before a finite order parameter is established.  The relation between the two phenomena has not been discussed elsewhere although the symmetry requirements are very similar. In this section, we calculate the paraconductivity of the chiral nanotubes described by Eq. (\ref{eq:F2}) and discuss it in the same framework as we discuss the SDE. 

\begin{figure}
	\includegraphics[width=0.48\textwidth]{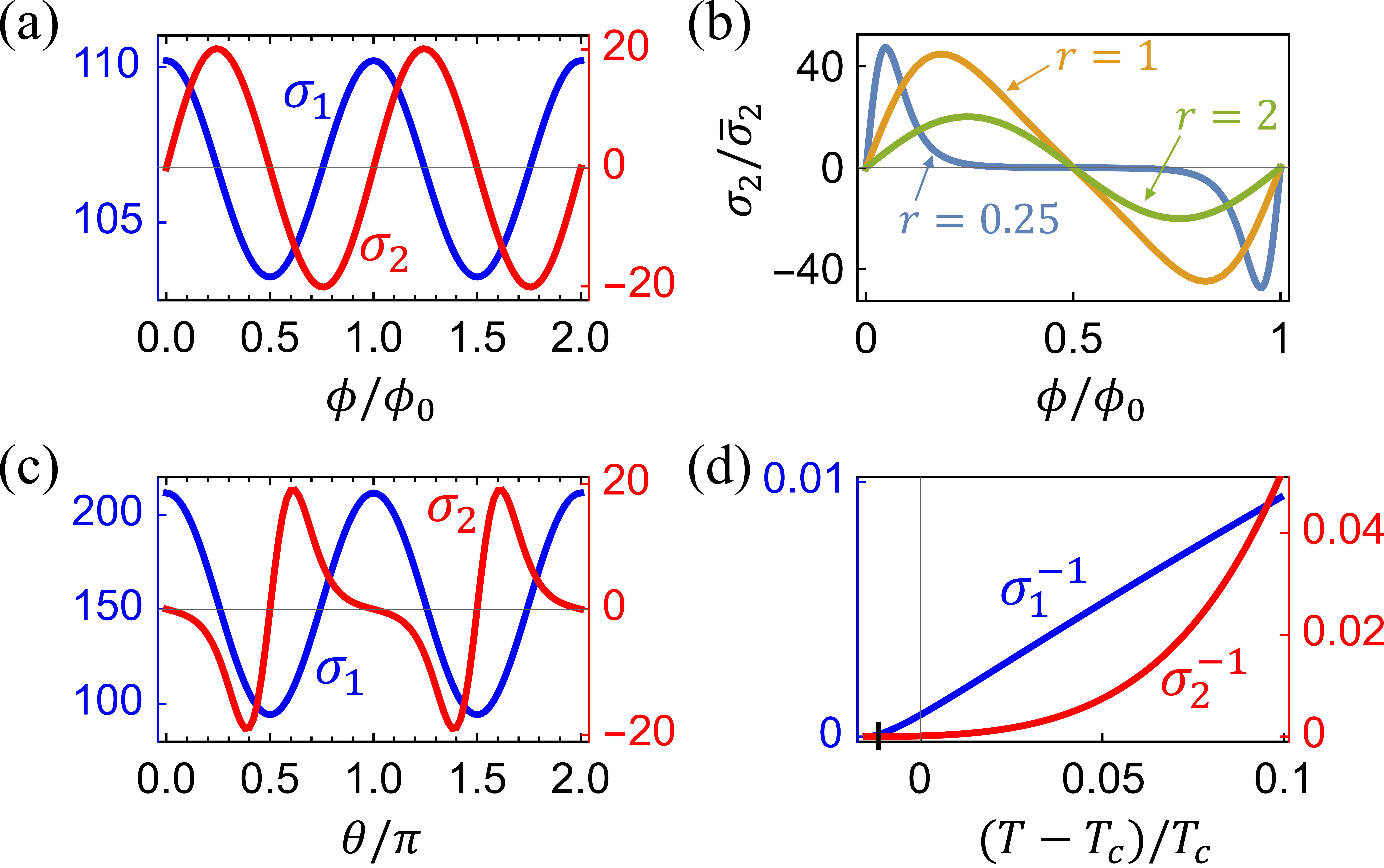}
	\caption{The linear and nonlinear paraconductivity of a chiral nanotube, $ \sigma_1 $ and $ \sigma_2$,  normalized by $ \bar{\sigma}_1 =  \frac{k_B T}{T_c } \frac{e^2}{4\pi^2 \hbar } \frac{\gamma l_0}{R } $ and $ \bar{\sigma}_2 =  \frac{k_B T}{T_c^2 } \frac{e^3}{6\pi^2 \hbar }\frac{\gamma^2 l_0^2}{R}$ respectively. 
	(a) Magnetic-flux dependence, showing the Little-Parks oscillation. (b) The evolution of the flux dependence as the normalized radius $ r $ is varied. (c) Dependence of $ \sigma_2 $ on the chiral angle $ \theta $. (d) The temperature dependence of the inverse of $ \sigma_{1/2}$. The parameters are the same as those in FIG. \ref{fig:eta}.}
	\label{fig:sigma}
\end{figure}

We calculate the paraconductivity using the time-dependent Ginzburg-Landau theory \cite{Schmid} (see Materials and Methods). The resulting current density $ j_x = \sigma_1 E + \sigma_2 E^2 + O(E^3) $  where the linear conductivity
\begin{align}
	\sigma_1 = \gamma \frac{T}{T_c } \frac{e^2}{4\pi^2 \hbar } \frac{l_0}{R} \sum_n \int dx \frac{\partial^2_x f_n(x) }{[\alpha/T_c+f_n(x)]^2},
	\label{eq:sigma1}
\end{align}
and the nonreciprocal term
\begin{align}
	\sigma_2 = \gamma^2 \frac{T}{T_c^2 } \frac{e^3}{6\pi^2 \hbar }\frac{l_0^2}{R} \sum_n \int dx \frac{\partial^3_x f_n(x) }{[\alpha/T_c+f_n(x)]^3}	.
	\label{eq:sigma2}
\end{align}
In the dimensionless function $ f_n(x) =\xi(\bm p)/ T_c  $, we made a change of variables, $ \bm p =[p_x,p_y] \rightarrow [x \hbar q_0, y_n \hbar q_0 ] $, where $ y_n= (n-\phi/\phi_0)/Rq_0 $ and $ q_0  = 1/l_0 $. The substitution of Eq. (\ref{eq:xi}) leads to
\begin{align}
	 f_n(x) 
	 &=\frac{1}{T_c} \xi \left( x \hbar q_0, y_n\hbar q_0 \right) \notag\\ 
	 &= \frac{x^2}{2\tilde{m}_x} + \frac{y_n^2}{2\tilde{m}_y} + \frac{x y_n}{\tilde{m}_{xy} } + \sum_{i=0}^4 \tilde{\kappa}_n x^i y_n^{4-i}
\end{align}
where $ \tilde{m}_{x/y/xy} =  m_0^{-1}m_{x/y/xy} $ and $ \tilde{\kappa}_i = \kappa_i m_0 ( \hbar q_0)^2 $ are dimensionless parameters.

The integrals in Eqs. (\ref{eq:sigma1}) and (\ref{eq:sigma2}) can be done numerically and the resulting $ \sigma_{1/2} $ are shown in FIG. \ref{fig:sigma} as functions of the magnetic flux $ \phi $ and the chiral angle $ \theta  $. 
Little-Parks oscillations of both the linear and nonlinear conductivities are found in FIG. \ref{fig:sigma}(a). The maxima/minima of $ \sigma_1 $ are at integer/half-integer values of $ \phi/\phi_0 $ since $ \sigma_1 $ is an even function of $ \phi $ and finite flux suppresses superconductivity. On the hand, the nonreciprocal $ \sigma_2 $ is odd in $ \phi $ and it  vanishes whenever $ \phi/\phi_0 $ becomes a integer. The flux values for optimal $ \sigma_2 $ depend on the system parameters such as the nanotube radius, as shown in FIG. \ref{fig:sigma}(b). The curves resemble those in FIG. \ref{fig:eta}(a) with the difference that they are smooth here because all the transverse components $ n \in (-\infty, \infty) $ of the order parameter contribute, unlike the supercurrent which is given by a certain $ n $. 
FIG. \ref{fig:sigma}(c) shows the effect of the chiral angle $ \theta $. The angle dependence of $ \sigma_2 $ is of similar amplitude to the flux dependence  in FIG. \ref{fig:sigma}(a). 
In FIG. \ref{fig:sigma}(d), we find that the temperature dependence of $ \sigma_1 $ is rather linear, which is similar to higher-dimensional systems \cite{Schmid, Wakatsuki1,Wakatsuki2}. A difference here is a shifted transition temperature $ T_c' $, so that $ \sigma_1^{-1} \sim (T-T_c' )$. The $ T $-dependence of $ \sigma_2^{-1} $ is clearly of higher order and we do not find any single power law.  
\\

\paragraph{\bf DISCUSSION\\}

We have shown that superconducting chiral nanotubes with trapped magnetic flux behave as supercurrent diodes, whose diode efficiency strongly depends on the chiral angle. We also found, in the same theoretical framework, that the paraconductivity of such chiral nanotubes near $ T_c $ contains a nonreciprocal part $ \sigma_2 $, whose dependence on the system parameters is rather similar to that of the SDE efficiency $ \eta $ and oscillates periodically as the magnetic flux $ \phi $ or the chiral angle $ \theta $ is varied. The results show that a combination of inversion symmetry breaking by chiral structure and time-reversal symmetry breaking by magnetic flux can induce nonreciprocal transport properties, including the SDE and the NPC, in superconductors. 

One may notice that actual nanotubes created in laboratories are mostly related to honeycomb or triangular lattices while the nanotubes discussed here are obtained by rolling a sheet of rectangular lattice. This choice is for technical convenience.  However, the main conclusions drew here shall generally apply. 
To discuss a carbon nanotube (honeycomb) or a transition-metal-dichalcogenide nanotube (triangular), terms up to the 6-th order in momentum must be included when constructing their Ginzburg-Landau free energies, which is not really meaningful considering the condition for the validity of the Ginzburg-Landau theory itself. Thus, a study of realistic (carbon/NbSe2/...) nanotubes may need to use the microscopic BCS theory, which can be done numerically. 

Although single superconductors are considered here, the nonreciprocal effects discussed here shall apply to Josephson junctions where two conventional bulk superconductors (Al, Pb, Nb, NbSe2, etc.) are connected by a chiral nanotube. A study of such a system will be of great practical importance. In this manuscript, we aim to clarify the physical principles and general features of the nonreciprocal properties of superconducting chiral nanotubes, and leave more detailed and realistic studies to future works. 

Although one needs to break both $ \mathcal{P} $ and $ \mathcal{T} $ to obtain unequal $ J_{c\pm} $ \cite{JiangK,CJWu}, it should be noted that there also exist nonreciprocal properties in $ \mathcal{T} $-preserving Josephson junctions. The nonreciprocity may be observed in unequal retrapping currents $ J_{r\pm} $ \cite{Misaki} or in ac Josephson effects \cite{JPHu,JiangK}. The interaction between electrons plays an important role in these cases. The design or improvement of supercurrent diodes with strong electron interactions is a topic worth further investigation. 
\\

\paragraph{\bf MATERIALS AND METHODS\\}
\paragraph{\bf Parameters in the rotated coordinate system \\}
By rotating the coordinate system by the chiral angle $ \theta $, one obtains the free energy form in Eq. (\ref{eq:xi}) where the parameters are functions of those in Eq. (\ref{eq:xi0}). The functional forms are
\begin{align}
	&\frac{1}{m_{x/y}} = \frac{1}{m_0} \pm \frac{\cos 2\theta }{m_1},\label{eq:mx}\\
	&\frac{1}{m_{xy}} =  -\frac{\sin 2\theta }{m_1},\label{eq:mxy}\\
	&\kappa_0 = \kappa_4  = \frac{1}{4} \left(\frac{1}{\zeta_0 m_0^2}+\frac{\cos ^2 2 \theta }{\zeta_1 m_1^2}+\frac{\cos 4 \theta }{\zeta _2 m_2^2}\right), \\
	&\kappa_1=-\kappa_3 = \frac{ \sin 4 \theta }{2} \left(\frac{1}{\zeta _1 m_1^2}+\frac{2}{\zeta _2 m_2^2}\right),\label{eq:kappa1}\\
	&\kappa_2 = \frac{1}{4} \left(\frac{2}{\zeta _0 m_0^2}+\frac{1-3 \cos 4 \theta }{\zeta _1 m_1^2}-\frac{6 \cos 4 \theta }{\zeta _2 m_2^2}\right).
	\label{eq:kappa2}
\end{align}
\paragraph{\bf Time-dependent Ginzburg-Landau theory\\}
At a temperature slightly above $ T_c $, the fluctuation of the order parameter is determined by the following Langevin equation \cite{Schmid},
\begin{align}
	\hbar\gamma \partial_t \psi(\bm r,t) = -\left[\alpha   + \xi(\hat{\bm p}) \right]\psi(\bm r,t) + \delta(\bm r,t),
	\label{eq:TDGL}
\end{align}
where $\delta(\bm r,t) $ is an uncorrelated random force and  $ \gamma $ is the inverse of damping constant. 
Note that $ \alpha>0 $ and the static order parameter vanishes, i.e. $ \langle \psi_{n,q}(t) \rangle_t =0$. However, Eq. (\ref{eq:TDGL}) leads to a nonzero $ \langle |\psi_{n,q}(t)|^2 \rangle_t $, which is \cite{Schmid}
\begin{align}
	\langle |\psi_{n,q}(t)|^2 \rangle = \frac{2k_B T}{\hbar \gamma } \int_{-\infty}^t dt' e^{ -\frac{2}{\hbar \gamma} \int_{t'}^{t} dt'' [\alpha + \xi(t'')]  }. 
	\label{eq:mean}
\end{align}
It is nonzero when an electric field $ \bm E = E \hat{\bm x} $ is applied, making $ \xi(\bm p(t'')) = \xi_n(q + 2eEt'') $. Combining Eqs. (\ref{eq:Jx0}) and (\ref{eq:mean}), one obtains Eq. (\ref{eq:sigma1}) and Eq. (\ref{eq:sigma2}).

\bibliographystyle{apsrev4-1}

\begin{acknowledgments}

	\paragraph{\\ \bf Acknowledgments\\}	
	{\bf Funding:}	N.N. was supported by JST CREST Grant Number JPMJCR1874, Japan, and JSPS KAKENHI Grant Number 18H03676. Y.T. was supported by Scientific Research (A) (KAKENHI Grant No. JP20H00131), Scientific Research (B) (KAKENHI Grants No. JP20H01857) and JSPS Core-to-Core program Oxide Superspin international network (Grants No. JPJSCCA20170002).
	{\bf Author contributions:} N.N. initiated and supervised this work. Y.T. helped to analyze the problem. J.J.H. carried out the calculations and wrote the manuscript with suggestions from all the authors. 
	{\bf Competing interests:}	The authors declare that they have no competing interests.
	{\bf Data and materials availability: }	All data needed to evaluate the conclusions in the paper are present in the paper.

\end{acknowledgments}

\end{document}